\documentstyle[psfig,twoside,fleqn,espcrc2]{article}
\title{ $J/\psi$ Production at the LHC} 
\author{Miguel     Angel
Sanchis-Lozano
\thanks{PPE Division, CERN, CH-1211 Geneva 23, Switzerland. 
E-mail: mas@evalvx.ific.uv.es} and Beatriz Cano-Coloma \thanks{IFIC,
46100 Burjassot, Valencia, Spain. 
E-mail: bcano@evalo1.ific.uv.es}} 

\begin{document}

\begin{abstract}
  We firstly examine  hadroproduction of  prompt  $J/\psi$'s at the  Fermilab
  Tevatron in a Monte Carlo framework by means  of the event generator PYTHIA
  5.7   in which    those  colour-octet processes    relevant  for charmonium
  production have  been  implemented accordingly.  We find that  colour-octet
  matrix  elements presented in literature  from $p\bar{p}$ collider data are
  {\em  systematically}  overestimated  due  to    overlooking of   the  {\em
  effective} primordial   transverse  momentum of  partons  (i.e.   including
  higher-order QCD effects).   We estimate the   size of these  effects using
  different parton  distribution functions.  Finally, after normalization  to
  Tevatron data, we extrapolate up to LHC energies making a prediction on the
  expected $p_t$ differential cross-section for charmonium. 
\end{abstract}

\maketitle

The experimental discovery at Fermilab \cite{fermi} of an excess of inclusive
production  of prompt heavy  quarkonia (mainly  for  $J/\psi$ and  $\psi(2S)$
resonances) in  antiproton-proton collisions triggered an intense theoretical
activity beyond what was considered  conventional  wisdom just few years  ago
\cite{greco}. Indeed, the   discrepancy between the  so-called colour-singlet
model (CSM) in hadroproduction \cite{baier} and the experimental data amounts
to more  than an order   of magnitude  and   cannot  be attributed to   those
theoretical uncertainties  arising from the  ambiguities on  the choice of  a
particular    parton distribution function (PDF),    the  heavy quark mass or
different energy scales. 

Recently it  has been argued that the  surplus of charmonia production can be
accounted for by assuming that the heavy quark pair not necessarily has to be
produced in a colour-singlet  state  at the short-distance partonic   process
itself \cite{braaten}.   Conversely,  it can  be  produced in a  colour-octet
state evolving non-perturbatively into  quarkonium in a specific  final state
with the correct quantum  numbers according to some computable  probabilities
governed by the internal velocity of the heavy quark. This mechanism, usually
named as   the colour-octet  model  (COM),   can be  cast into  the  rigorous
framework of an effective non-relativistic theory for the strong interactions
(NRQCD) deriving from first principles \cite{bodwin}.

However, the weakness  of the COM lies  in the fact that the non-perturbative
parameters characterizing  the long-distance hadronization process beyond the
colour-singlet contribution  (i.e. the colour-octet  matrix elements)  are so
far almost free parameters to be adjusted  from the fit to experimental data,
though expected to be  mutually   consistent according  to the  NRQCD   power
counting rules. 

On  the other  hand, an  attractive   feature of the colour-octet  hypothesis
consists of the  universality of the  NRQCD matrix elements entering in other
charmonium production processes.  Let us look below in some detail at the way
hadronization is  folded with the partonic  description of  hadrons, focusing
for concreteness on the couple of related papers \cite{cho0} \cite{cho}. 

In the first paper \cite{cho0},  Cho and Leibovich considered for  charmonium
production only the $^3S_1$ coloured intermediate state  as a first approach,
computing the  matrix elements  as products  of  perturbative  parts  for the
short-distance   partonic  processes, and   the  colour-octet  matrix element
concerning the long-distance hadronization, in  a similar but generalized way
as   in the CSM.   Finally, a  convolution   of concrete  parton distribution
functions and the   differential cross-section for the $Q\bar{Q}$  production
subprocess was performed, whereby    the $p_t$ dependence of   the  charmonia
production coming from  the latter. Moreover, in  ref.  \cite{cho} the same
authors take into account further contribution from  new coloured states (for
more details see    the quoted references) concluding  finally   that at high
enough $p_t$ a two-parameter fit is actually required to explain the observed
inclusive  $p_t$   distribution  of charmonia   production  at  the Tevatron.

However, it  is  well-known for a long   time that higher-order effects  ($K$
factors)    play    an   important     role  in  inclusive    hadroproduction
\cite{break}. Namely,   beyond the primordial  transverse   momentum $k_t$ of
partons in hadrons, higher-order QCD effects  such as initial-state radiation
of gluons  by the  interacting partons  add up to   yield an  {\em effective}
intrinsic transverse momentum which  certainly has to  be considered in charm
production even at  high $p_t$ \cite{leo}. As  we shall see, if overlooked at
all, the effect  on the fit  parameters  (and ultimately on  the colour-octet
matrix elements) amounts to a {\em systematic} overestimate. 

\begin{figure}[htb]
\psfig{figure=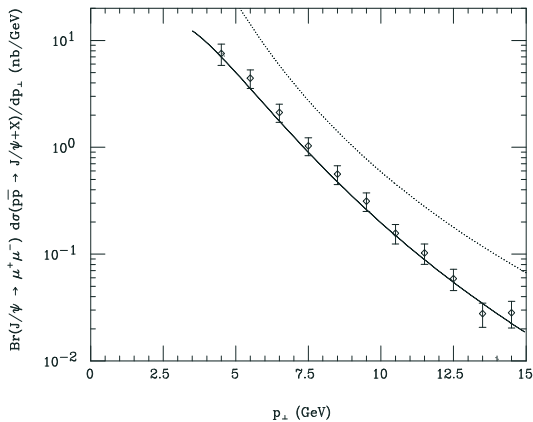,height=5.5cm,width=7.cm}
\caption{ Curves obtained from PYTHIA (not fit) including colour-octet
mechanism for prompt $J/\psi$ production at the Tevatron using the same
parameters as in Ref. [7]. The
solid line corresponds to initial- and final-state radiation
turned off and the dotted line to  initial- and
final-state radiation on. The MRSD0 parton distribution
function was employed as in  [7].} 
\end{figure}

In   this  work we  have   implemented  in  the  event   generator PYTHIA 5.7
\cite{pythia} those new colour-octet mechanisms  for the dominant gluon-gluon
fusion  process \footnote{Originally  PYTHIA generates   direct production of
$J/\psi$'s via the CSM \cite{pythia}. Technically, we have implemented in the
PYTHIA subroutine PYSIGH  the two extra  differential cross-sections for  the
process $g+g{\rightarrow}J/\psi+g$  corresponding to the coloured $^3S_1$ and
$^1S_0+^3P_0$  states     whose   matrix  elements    can     be  found    in
\cite{cho0}\cite{cho}.}.  Let  us  stress that in our  analysis initial-state
radiation was  incorporated  within the framework  of  the  Lund Monte  Carlo
framework \cite{sjos}   \footnote{Initial-state radiation is   switched on by
default in PYTHIA: MSTP(61) is the master switch \cite{pythia}.}. 

\begin{figure}[htb]
\psfig{figure=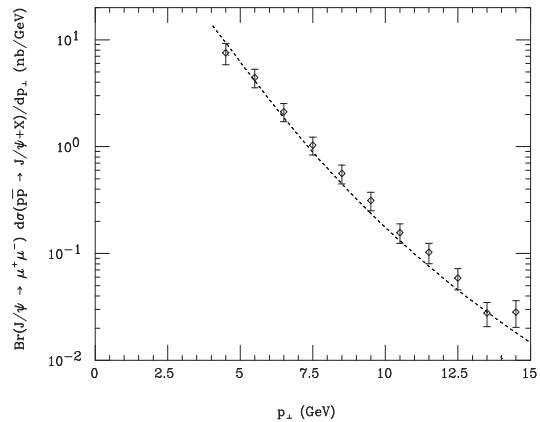,height=5.5cm,width=7.cm}
\caption{Two-parameter fit to the experimental Tevatron data, where initial-
and final-state radiation were incorporated via PYTHIA generation. The CTEQ
2L parton distribution function was employed in this plot.
} 
\end{figure}

\begin{figure}[htb]
\psfig{figure=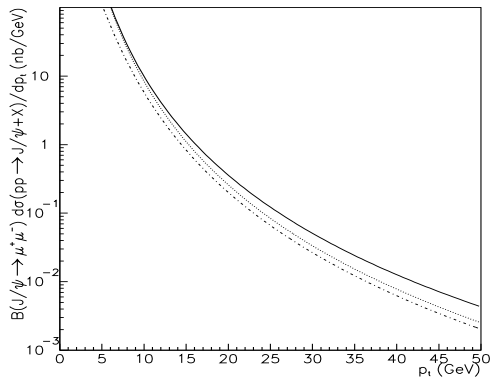,height=5.7cm,width=7.cm}
\caption{Our prediction for prompt $J/\psi$ 
direct production
at the LHC using PYTHIA with the new colour-octet matrix
elements 
 a) dotted line: CTEQ 2L, b) solid line: GRV HO, and
c) dot-dashed line: MRSD0. The rapidity cut 
${\mid}y{\mid}<2.5$ on the $J/\psi$ was required.
} 
\end{figure}

\begin{table*}[hbt]
\setlength{\tabcolsep}{1.5pc}
\newlength{\digitwidth} \settowidth{\digitwidth}{\rm 0}
\caption{Colour-octet matrix
elements (in units of GeV$^3$) from the fit
of Tevatron data on prompt $J/\psi$ production
for different parton distribution functions. For comparison we
quote the values given in Ref. [7]: 
$(6.6{\pm}2.1){\times}10^{-3}$ and $(2.2{\pm}0.5){\times}10^{-2}$  respectively.}
\label{FACTORES}

\begin{center}
\begin{tabular}{lcc}    \hline
matrix element  & $<0{\mid}O_8^{J/\psi}(^3S_1){\mid}0>$ & 
$\frac{<0{\mid}O_8^{J/\psi}(^3P_0){\mid}0>}{M_c^2}+
\frac{<0{\mid}O_8^{J/\psi}(^1S_0){\mid}0>}{3}$ \\
\hline
CTEQ2L &$3.3{\times}10^{-3}$& $1.1{\times}10^{-2}$ \\
MRSD0 & $4.3{\times}10^{-3}$ & $4.4{\times}10^{-3}$ \\
GRV 94 HO & $4.6{\times}10^{-3}$ & $3.5{\times}10^{-3}$ \\
\hline
\end{tabular}
\end{center}
\end{table*}

Using  the  same numerical  values  for the  colour-octet matrix  elements as
reported in table I of Ref.  \cite{cho} if initial- and final-state radiation
are  turned off, a relative  good agreement of the  theoretical curve and the
experimental points  is  obtained  (see Fig.  1),   as  should be  reasonably
expected   \footnote{The same PDF  as in  Ref.   \cite{cho} (MRSD0) was used.
Still the theoretical curve  stands  slightly below the experimental   points
maybe due to the fact that we have neglected other non-dominant channels like
$q\bar{q}{\rightarrow}J/{\psi}g$  and  $qg{\rightarrow}J/{\psi}g$  \cite{cho}
over the parton $x$ range   $\simeq[10^{-3},10^{-2}]$. Nevertheless the  fact
that we   somehow  underestimate the  production  rate of    prompt charmonia
reinforces our arguments favouring  lower values for the  colour-octet matrix
elements.}.  However, if  the initial- and final-state radiation \footnote{It
should  be noted that  initial-state radiation and final-state radiation have
opposite  effects in the $p_t$ spectrum,  the former enhancing the high $p_t$
tail   whereas  the latter  smears  the  distribution.}  are switched  on the
predicted curve stands  well above  the  experimental data. By inspection  we
observe from  Fig.  1  that  the theoretical curve  including   initial-state
radiation  is shifted upwards with respect  to the experimental distribution,
in accordance   with  the  expected  ${\lq}{\lq}$kick"  caused  by the   {\em
effective} primordial transverse momentum of partons \cite{break} \cite{leo}.
Accordingly, keeping radiation effects  it turns out that  the values for the
colour-octet  matrix elements have   to  be  lowered by  significant  factors
(!). Of course, the PYTHIA treatment of the effective $k_t$ is not guaranteed
to be  perfect but, nevertheless, should  give a reasonable estimate  of such
effects.  

In  order  to assess   the importance  of  the effective intrinsic
transverse momentum  of partons we have made  three different choices for the
proton PDF \footnote{See    \cite{pdflib} for  technical  details about   the
package of  Parton Density Functions  available at  the CERN Program Library.
References therein.}: 

\begin{description} 
 \item[a)] the leading order CTEQ 2L (by default in PYTHIA 5.7)  
 \item[b)] the next to leading order MRSD0 (the same as used in \cite{cho}) 
 \item[c)] the next to leading order GRV 94 HO 
\end{description}

As already mentioned  before, the  theoretical  curve of the  inclusive $p_t$
distribution of prompt  $J/\psi$'s   stands   in all cases  above    Tevatron
experimental  points if  the set of  parameters from  \cite{cho}  are blindly
employed in the PYTHIA generation with initial-state radiation on.  Motivated
by this systematic discrepancy, we performed new fits for the prompt $J/\psi$
direct   production  at   Tevatron    (feed-down  from  radiative decay    of
${\chi}_{cJ}$  resonances  was   experimentally removed).   The corresponding
colour-octet matrix elements are shown in table I and the plot for case a) in
Fig.  2.

Finally,  we  have  generated  prompt $J/\psi$'s in   proton-proton
collisions at LHC  energies (center-of-mass energy = 14  TeV) by means of our
${\lq}{\lq}$modified" version of PYTHIA with the colour-octet matrix elements
as shown in Table I -  i.e. after normalization to  Tevatron data. Indeed, an
${\lq}{\lq}$order  of  magnitude"  estimate  of  the  expected  production of
charmonia at the  LHC  is suitable from   many points of  views \cite{atlas}:
$J/\psi$ can be considered either as a signal in its own right or as a source
of background for other  interesting processes involving $J/\psi$'s, like  CP
studies from  the cascade channel $B_d^0\  {\rightarrow}\ J/\psi\ K_s^0$.  In
Fig. 3 we show the $p_t$ inclusive distributions for direct prompt $J/\psi$'s
at the LHC obtained for each  of the three  PDF's employed in our study. Such
predictions  should be  taken, however, with   a grain  of  salt since gluons
emitted either at the initial or at the final state  could contribute at high
energies   to   the   final    yield  of  charmonia    via   the colour-octet
mechanism. (This possibility  is currently being investigated.)

In summary,
we have investigated  higher-order effects induced by initial-state radiation
on the extraction  of the NRQCD matrix elements  from hadroproduction at high
$p_t$   by  means of   an event   generator  (PYTHIA 5.7)   with colour-octet
mechanisms   implemented in.  We   conclude  that   for different  PDF's  the
overlooking of the {\em effective}  primordial $k_t$ leads sytematically to a
significant  overestimate  of  the  colour parameters.  We   have derived new
colour-octet matrix elements for $J/\psi$ production from Tevatron data using
three  different sets  of  PDF's.

Our  conclusion  is that  a  more detailed
analysis is actually required to  extract NRQCD colour-octet matrix  elements
from  hadroproduction data in  a reliable way  thereby  allowing a meaningful
comparison with other charmonia production processes, like photoproduction at
HERA \cite{cacciari}. Finally we have estimated the prompt $J/\psi$ inclusive
direct  production at the  LHC for three different  PDF's, finding an overall
relative good accordance at moderate transverse momentum. 

\section*{Acknowledgments}

We thank P. Eerola and N. Ellis and the ATLAS B physics
working group for useful comments and an encouraging attitude. 
Comments by S. Baranov, E. Kovacs and L. Rossi are acknowledged. We also 
thank T. Sj\"{o}strand for his advice on some important aspects 
of our  work.
Research partially supported by CICYT under contract AEN-96/1718.

\end{document}